\documentclass
[11pt,tightenlines,eqsecnum,floats,aps,amsmath,amssymb,nofootinbib,prd,shownopacs,notitlepage]{revtex4}
\usepackage{graphicx, wrapfig}
\usepackage{amssymb}
\usepackage{color}
\usepackage{mathrsfs}
\usepackage{amsmath}
\usepackage{afterpage}
\usepackage{enumerate}
\usepackage{epstopdf}
\usepackage{ulem}
\DeclareGraphicsExtensions{.pdf,.png,.jpg,.jpeg,.eps}

\newcommand{\be}{\nopagebreak[3]\begin{equation}}
\newcommand{\ee}{\end{equation}}
\newcommand{\bfig}{\nopagebreak[3]\begin{figure}}
\newcommand{\efig}{\end{figure}}
\newcommand{\bea}{\nopagebreak[3]\begin{eqnarray}}
\newcommand{\eea}{\end{eqnarray}}

\newcommand{\bmult}{\nopagebreak[3]\begin{multline}}
\newcommand{\emult}{\end{multline}}

\begin{document}

\title{Anomalies in the Cosmic Microwave Background and their Non-Gaussian Origin in Loop Quantum Cosmology\footnote{Contribution to the Special Issue ``Loop Quantum Cosmology'' in Frontiers  in Astronomy and Space Sciences, edited by B. Elizaga-Navascu\'es, G. Mena-Marug\'an, and F. Vidotto.} }
\author{Ivan Agullo}
\email{agullo@lsu.edu}
\affiliation{Department of Physics and Astronomy, Louisiana State University, Baton Rouge, LA 70803, USA}
\author{Dimitrios Kranas}
\email{dkrana1@lsu.edu}
\affiliation{Department of Physics and Astronomy, Louisiana State University, Baton Rouge, LA 70803, USA}
\author{V. Sreenath}
\email{sreenath@nitk.edu.in}
\affiliation{Department of Physics, National Institute of Technology Karnataka, Surathkal, Mangalore 575025, India}

\begin{abstract}

Anomalies in the cosmic microwave background (CMB) refer to  features that have been observed, mostly  at large angular scales, and which show some  tension with the statistical predictions of the standard $\Lambda$CDM model. In this work, we focus our attention on power suppression, dipolar modulation, a preference for odd parity, and the tension in the lensing parameter $A_L$. Though the statistical significance of each individual anomaly is inconclusive, collectively they are significant, and could indicate new physics beyond the $\Lambda$CDM model. In this article, we present a brief, but pedagogical introduction to CMB anomalies and  propose a common origin in the context of loop quantum cosmology. 
\end{abstract}

\maketitle
\section{Introduction}

Observations of the cosmic microwave background (CMB) by the Planck satellite have revealed that the $\Lambda$CDM model together with the inflationary scenario checks nearly all the right boxes \cite{Akrami:2018vks, Aghanim:2018eyx, Akrami:2018odb}---in the sense that  it provides a detailed fit to the CMB spectrum based on a few free parameters  \cite{Aghanim:2018eyx, Aghanim:2019ame}. The nearly scale invariant power spectrum predicted by slow roll inflation has been confirmed with a significance of more than $7  \sigma$ \cite{Aghanim:2018eyx, Akrami:2018odb}. Further, observations are consistent with the near Gaussian nature of the primordial perturbations predicted by slow roll inflation \cite{Akrami:2019izv}. 
\par
But in spite of  this success,  several open questions remain. A prominent one concerns the past incompleteness of the inflationary scenario. As it is well known, general relativity,  on which the inflationary scenario rests, breaks down as we approach the Planck regime. Loop quantum cosmology (LQC) is an attempt to use the principles of loop quantum gravity to address this issue \cite{Bojowald:2001xe,Ashtekar:2003hd,MenaMarugan:2011va,Banerjee:2011qu,Ashtekar:2011ni,Agullo:2016tjh,Agullo:2013dla}. 
In LQC, the big bang singularity is replaced by a {\it bounce} \cite{Ashtekar:2006rx,Ashtekar:2006wn}, which is triggered  by quantum gravitational effects.  This bounce by itself is not able to generate the primordial perturbations though, and it must be complemented with another mechanism. A natural strategy is to maintain the inflationary phase in the post-bounce era. In such scenario, the  goal of the bounce is, in addition to overcoming the difficulties arising from classical general relativity, to bring the universe to an inflationary phase.
From a practical viewpoint, the pre-inflationary bounce can be thought  as a mechanism to specify the initial conditions for scalar and tensor perturbations at the onset of inflation. Numerous studies have shown that the bounce predicted by LQC modifies the primordial power spectra of scalar and tensor perturbation \cite{Bojowald:2008jv,Bojowald:2010me,Agullo:2012sh,Agullo:2012fc,Agullo:2013ai, Fernandez-Mendez:2013jqa,Fernandez-Mendez:2014raa,Barrau:2014maa,deBlas:2016puz,Agullo:2015tca,Agullo:2016hap,Ashtekar:2016wpi,Martinez:2016hmn, Gomar:2017yww,Zhu:2017jew,Agullo:2018wbf,Li:2019qzr,ElizagaNavascues:2020fai,Li:2020mfi,Agullo:2020wur,Agullo:2020iqv,Ashtekar:2020gec,ElizagaNavascues:2020uyf,Martin-Benito:2021szh} and the non-Gaussianity \cite{Agullo:2017eyh,Agullo:2015aba,Sreenath:2019uuo,Zhu:2017onp} at large angular scales, while at smaller scales in the CMB the predictions are indistinguishable from those of standard inflation with Bunch-Davies initial conditions. 
Hence, if at all early universe scenarios such as LQC were to leave any imprints on the CMB, they would be expected at the longest observable scales, or equivalently, at the lowest  angular multipoles.
\par
It is for this reason  that certain puzzling signatures  which  have been recently observed at  large angular scales in the CMB become relevant \cite{Akrami:2019bkn}. These signatures, generically known as CMB anomalies, are features  that are in conflict with the almost scale invariance predicted by inflation, or with the  statistical isotropy and homogeneity assumed in the $\Lambda$CDM.  In more detail, the anomalies observed by Planck include a lack of two-point correlations at large angular scales, a dipolar asymmetry, a preference for odd parity, alignment of low multipoles, a cold spot, etc. In addition, the Planck analysis  has also found a preference for a larger value of the lensing parameter \cite{Aghanim:2018eyx} than it is expected. Some of these anomalies were already  observed by the WMAP satellite and even by COBE. Hence, the consensus is that  these signals are not due to unaccounted systematics. Put it simply, there is no debate about the fact that these are real features in the CMB (see e.g. \cite{Schwarz:2015cma}).  However, the statistical significance with which these features depart from the predictions of  the $\Lambda$CDM model is, though non-negligible, inconclusive, and the debate is rather whether any of these features are significant enough to require the introduction of new physics. Recall that the $\Lambda$CDM only makes statistical predictions, and therefore none of these features are actually incompatible with $\Lambda$CDM. But if we accept the $\Lambda$CDM model, they imply that we happen to live in an uncommon realization of the underlying probability distribution. Another possibility is that some or all these features are  signatures of new physics, and they are in fact expected signals in a suitable extension of the $\Lambda$CDM theory. %
\par

In  recent work \cite{Agullo:2020fbw,Agullo:2020cvg} we proposed that a cosmic bounce before  inflation naturally  changes the primordial probability distribution in such a way that, in a statistical sense, the observed features are not anomalous.
The core of the idea is that a cosmic bounce generates strong  correlations (non-Gaussianites) between the longest modes we observe in the sky and longer, super-horizon modes. We cannot observe directly these correlations since some of the modes involved have wave-lengths larger than the Hubble radius today. But these correlations produce indirect effects in observable modes, which can account for the observed  anomalies. The goal of this article is to apply the general ideas introduced in \cite{Agullo:2020fbw,Agullo:2020cvg} to LQC. We will also take the opportunity to provide a succinct and pedagogical introduction to CMB anomalies and the phenomenon of non-Gaussian modulation, addressed to the quantum cosmology community. See \cite{Agullo:2015tca, deBlas:2016puz,Ashtekar:2016wpi,Ashtekar:2020gec,Agullo:2020wur,Agullo:2020iqv} for other ideas to account for some of the features observed  in the CMB within LQC. In particular, the companion article \cite{Ashtekar:2021izi}  in this  special issue, provides an interesting set of complementary ideas and perspectives on  the way LQC can account for the CMB anomalies. 
\par
The plan of this article is as follows. In the next section, we discuss  the basic  principles behind quantifying temperature anisotropy and discuss the implications of statistical homogeneity and isotropy for CMB anisotropies. Then we describe  some of the anomalies observed by the Planck satellite, which point to a violation of the underlying assumption of statistical homogeneity and isotropy. In section \ref{sec:3}, we describe the mechanism behind the phenomenon of non-Gaussian modulation. In section \ref{sec:4}, we provide a quick description of the evolution of perturbations in LQC and discuss the power spectrum and bispectrum generated therein. We then apply non-Gaussian modulation to LQC in section \ref{sec:5} and present our results. In this section, we describe how the presence of non-Gaussian modulation in LQC makes these anomalous features more likely  to occur, in a way that they are no longer anomalous. Finally, in section \ref{sec:6}, we conclude with a discussion of our results, its short comings and future directions.

\section{Introduction to CMB anomalies}\label{sec:2}
The temperature $T(\hat n)$ of the CMB as a function of the direction $\hat n$ is nearly uniform, making it convenient to split $T(\hat n)$ into an isotropic part, the mean temperature $\bar T\, =\, \frac{1}{4\,\pi}\int {\rm d}\Omega \, T(\hat n)$, and the anisotropic deviation from it
\begin{equation}
  \delta T(\hat n)\, \equiv \,  \frac{T(\hat n)\, -\, \bar T}{\bar T}\, =\, \sum_{\ell\,m} a_{\ell\,m}\,Y_{\ell\, m}(\hat n), 
\end{equation}
where in the last equality we have decomposed the function $\delta T(\hat n)$ in spherical harmonics 
 $Y_{\ell m}$.  (see, for instance, \cite{Weinberg_2008, Durrer:2008eom}).
The mean temperature $\bar T$ is a free parameter of the $\Lambda CDM$ model, which is determined by observations. Our best measurement of $\bar T$ comes from the FIRAS instrument in the COBE satellite, and  is measured $\bar T\, =\, 2.725\, \pm\, 0.002\, K$ \cite{Fixsen:1996nj}.
\par

The $\Lambda CDM$ model predicts only the statistical properties of the temperature map $\delta T(\hat n)$ or, equivalently, of the coefficients $a_{\ell \, m}$. Therefore, the quantities we want to extract from observations are the moments:  $\langle a_{\ell \, m}a_{\ell' \, m'}\rangle$,  $\langle a_{\ell \, m}a_{\ell' \, m'}a_{\ell'' \, m''}\rangle$, etc. There are theoretical reasons, further supported by observations, to argue that the probability distribution we are after is very close to Gaussian, in which case the simplest non-zero moment, $\langle a_{\ell \, m}a_{\ell' \, m'}\rangle$, is all we need (recall that a Gaussian distribution is completely characterized by the mean and the variance). Furthermore, the assumption of statistical homogeneity and isotropy, on which the $\Lambda$CDM model rests, implies that $\langle a_{\ell \, m}a_{\ell' \, m'}^*\rangle$ must be diagonal in $\ell$ and $m$, and $m$-independent

\begin{equation}
    \langle a_{\ell m}\, a_{\ell' m'}^*\rangle\, = \, C_{\ell}\, \delta_{\ell \ell'}\, \delta_{m m'}.
\end{equation}
In other words, homogeneity and isotropy  imply that all information contained in the second moments can be codified in the $m$-independent coefficients $C_{\ell}$, for $\ell=1,2,3,\dots$. $C_{\ell}$ is known as the angular power spectrum. 

The equivalent statement in angular space is that the second moments of $\delta T(\hat n)$, $C(\theta)\, \equiv\, \langle \delta T(\hat n)\, \delta T(\hat n') \rangle$ can only depend on the angle $\theta$ between the two directions $\hat n$ and $\hat n'$:
\begin{equation}
    C(\theta)\, \equiv \, \langle \delta T(\hat n)\, \delta T(\hat n') \rangle\, =\, \frac{1}{4\pi}\, \sum_\ell (2\,\ell\, +\, 1)\, C_\ell\, P_\ell (\cos \theta) \, .
\end{equation}
If the assumptions of statistical homogeneity and isotropy break down, then the simple characterization  of the two-point correlations in terms of the simple quantity $C_{\ell}$ or $C(\theta)$ becomes  insufficient, and one would have to work with the full covariance matrix of $a_{\ell\,m}$ or $\delta T(\hat n)$. 

 The angular power spectrum $C_{\ell}$ is measured by averaging the data from satellites.  But, what is the correct notion of average? Ideally, one would like to have different realizations of the probability distribution (that is, different universes) and take averages on them, which is closer to the way averages are measured in quantum systems. Another possibility is to take averages over the CMB temperature map observed from different locations in the universe. The ergodic theorem relates both averages.  Unfortunately, none of these two strategies are available at the practical level.  Rather, what is  done in practice is to take advantage of the $m$-independence of the power spectrum $C_{\ell}$, and obtain it by averaging over its value obtained from individual $m$'s (we actually observe $\delta T(\hat n)$, but a simple computer code can translate the data to values of $a_{\ell\, m}$). The limitation of this strategy is clear: we have $2 \ell+1$ values of $m$ for each multipole $\ell$, and consequently the  uncertainty about the value of $C_{\ell}$ obtained in this way will be large for small values of $\ell$. This uncertainty is known as cosmic variance,  and it is quantified by $ \pm \sqrt{2/(2\ell + 1)}\, C_{\ell}$. It is not difficult to translate this uncertainty to angular space, and the result is $\pm \sigma(C(\theta))$, with
   \begin{equation}
    \sigma^2(C(\theta))\, =\, \frac{1}{8\,\pi^2}\, \sum_\ell\, (2\,\ell\, +\, 1)\,C_\ell^2\, P_\ell^2(\cos\theta).
\end{equation}   
Cosmic variance is an intrinsic limitation of cosmological observations, and cannot be overcome by building more precise instruments. Therefore, in making predictions for  $C_\ell$ or $C(\theta)$, one needs to keep in mind this inherent uncertainty. 
\par
\begin{figure}
    \centering
    \includegraphics[width=0.8\textwidth]{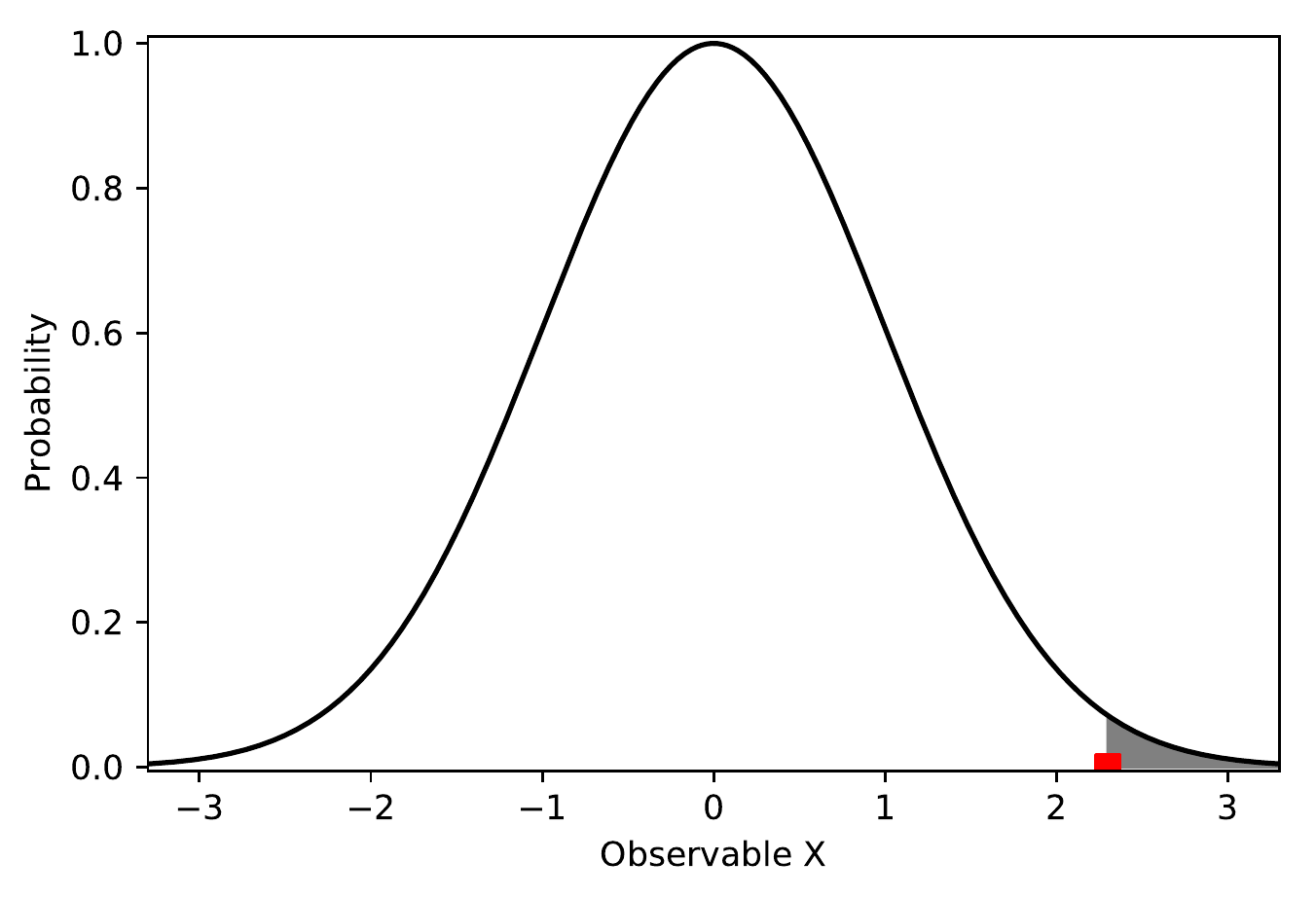}
    \caption{Illustration of the concept of $p$-value. The figure shows the  probability distribution of a certain observable $X$ according to the null hypothesis in black. The value of $X$ that is actually observed  is shown in red. Although the expected value of $X$ is zero, the observation is not incompatible with the theoretical prediction, given the statistical character of the later. The shaded area gives us the $p$-value of the observed value of $X$. As it  is evident from the figure, a smaller  p-value implies a larger  departure from the null hypothesis.}
    \label{fig:p-value}
\end{figure}

We now discuss the anomalous features that have been observed in CMB. The Planck team has carried out several tests to check the statistical isotropy of the CMB \cite{Ade:2013nlj, Ade:2015hxq, Akrami:2019bkn}. The CMB is a spherical shell of radiation, which captures a spherical sample of the density perturbations at the time of decoupling in the early universe. Hence, deviations from isotropy in the CMB sphere will signal deviation from statistical homogeneity or  isotropy in the early universe. Since, as emphasized above, the predictions from the $\Lambda$CDM model are statistical, a key aspect of the analysis is to quantify  the statistical significance of any observed departure from the theory. In statistical parlance, this is known as hypothesis testing, wherein a null hypothesis, which in this case is the $\Lambda$CDM model, is compared with observations. The departure from the null hypothesis is often quantified in terms of so called $p$-value. 
Given a null hypothesis, the $p$-value is the probability with which a certain phenomenon can occur. If the $p$-value of an observed feature is zero, the null hypothesis is automatically considered as incorrect. A very small value of the $p$-value, would  rather rule out the hypothesis with a  statistical significance given by $1-p$.  The concept  is visually illustrated in Fig. \ref{fig:p-value}: the $p$-value corresponds to the area of the shaded region. 
\par
In order to quantify an anomaly, the first step is to choose an observable of interest, which will serve as the indicator of the anomaly. Rather than analytically deriving the probability distribution  of the chosen observable out of the theory, a task that may be difficult for some observables,  in practice it is often  more convenient to estimate the $p$-value numerically. This can be done by  simulating a large number of random realizations of the CMB temperature map from the probability distribution of the $\Lambda$CDM model---using the best fit for the free parameters---and computing the $p$-value of the chosen observable from them. This is the way the Planck collaboration has evaluated the $p$-value of the anomalies discussed below \cite{Ade:2015via,Ade:2013nlj}. For example, if only five simulations out of a thousand lead to a value of an observable which is at least as extreme as the observed value, they would report a $p$-value of $0.005$  for that observation, or equivalently $0.5\%$. The anomalies considered in this article have $p$-value $\leq 1\%$ \cite{Schwarz:2015cma}. 
In the remaining part of this section, we briefly describe  the anomalies  that we consider in this article. 
\begin{figure}
    \centering
    \begin{tabular}{cc}
        \includegraphics[width=0.46\textwidth]{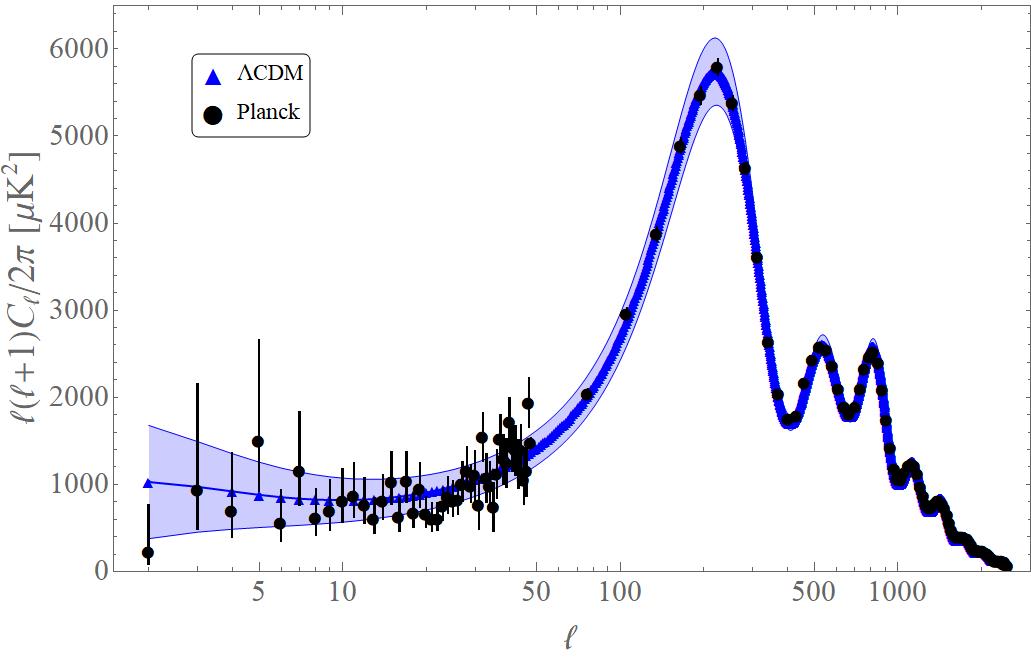}&
        \includegraphics[width=0.46\textwidth]{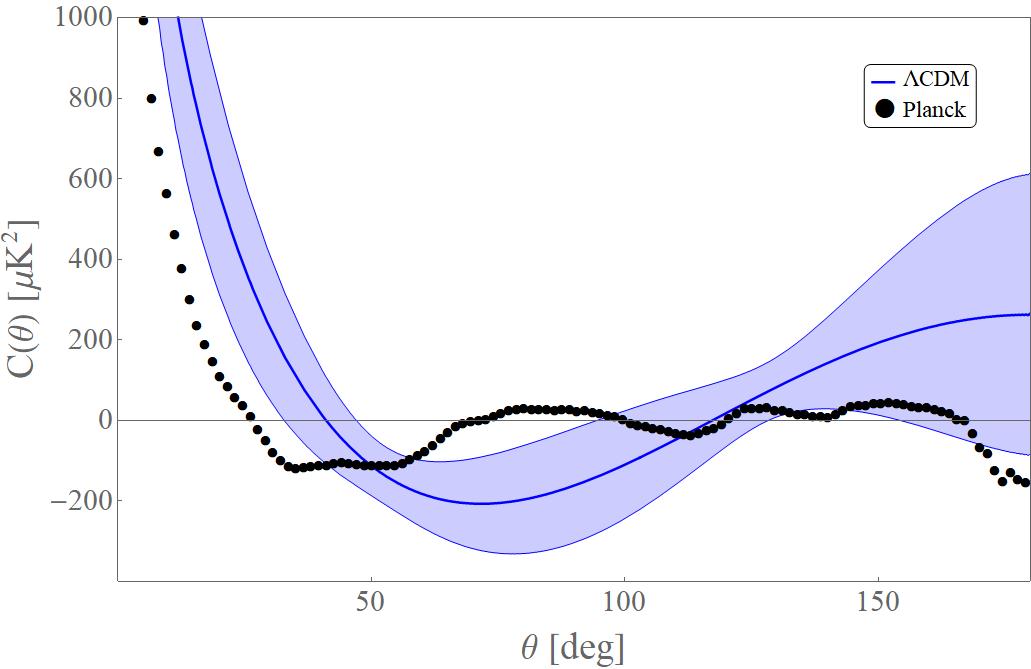}
    \end{tabular}
    \caption{The $TT$ power spectrum in the multipole (left) and angular space (right) generated in the standard model corresponding to the best fit parameters provided by Planck. The blue shaded region indicates the uncertainty due to cosmic variance. The black dots with error bars are the data from Planck. Note that the observed quadrupole is quite low compared to the prediction from the standard model, although it is compatible with the prediction within 1-$\sigma$ when we account for cosmic variance. The lack of power at large angular scales is more evident in the angular power spectrum, where the power is considerably low for angular scales greater than $60^{\circ}$. Furthermore, the amplitude of predicted power is larger than observed one by more than 1-$\sigma$ for the largest angular scales and for angles between $\approx\, 60^{\circ}$ to $80^{\circ}$. }
    \label{fig:power_suppression}
\end{figure}
\subsection {Power suppression}\label{suppression}
\par
Data from the satellites COBE \cite{Hinshaw:1996ut}, WMAP \cite{Bennett:2003bz} and Planck \cite{Akrami:2019bkn}, have consistently found a lack of two-point correlations at low multipoles, or at large angular scales, compared to what is expected in the $\Lambda$CDM model. Visually, this lack of correlations is evident in the real space two-point correlation function $C(\theta)$, shown in the right panel of Fig.~\ref{fig:power_suppression}: for angles larger than $60^{\circ}$, the two-point function is surprising low. The WMAP team had come up with an appropriate observable  to quantify this lack of power \cite{Spergel:2003cb}. It is defined by 
\begin{equation}
    S_{1/2}\, =\, \int_{-1}^{1/2}\, C(\theta)^2\, {\rm d}(\cos\, \theta)) \, .
\end{equation}
Its physical meaning is obvious: it captures the total amount of correlations squared (to avoid cancellations between positive and negative values of  $C(\theta)$) in angles $\theta\, >\, 60^0$. The $\Lambda$CDM model predicts $S_{1/2}\, \approx\, 42000 \mu K^4$, while the Planck satellite has reported a measured value\footnote{The value of $S_{1/2}$ varies a bit depending on the choice of map and the mask used.} of $S_{1/2}\, =\, 1209.2 \mu K^4$ \cite{Akrami:2019bkn}, which corresponds to a $p$-value less than $1\%$ ($\leq 0.5\%$  according to \cite{Schwarz:2015cma}). Put in simpler terms, this $p$-value tells us that if we were able to observe one thousand  universes ruled out by the $\Lambda$CDM model, only about a handful will show such a low value of $S_{1/2}$.

\subsection {Dipolar modulation anomaly}\label{dipolar}
A second important anomaly that has  been also observed by multiple satellites, is the presence of a dipolar modulation of the entire CMB signal \cite{Akrami:2019bkn}. 
This dipole should not be confused with the multipole $\ell=1$. Rather, the anomaly makes reference to {\it correlations} between multipoles $\ell$ and $\ell+1$, which can be explained by a {\it modulation} of  dipolar character, as we further discuss below. 

Such modulation was first modeled mathematically in \cite{Gordon_2005}, by adding a simple dipole to the temperature map as follows
\begin{equation}\label{eqn:gordon}
    T(\hat n)\, =\, T_0(\hat n)\, \left[ 1\, +\, A_1\,\hat n\cdot \hat d \right]\, ,
\end{equation}
where $T_0(\hat n)$ is the unmodulated (statistically isotropic) temperature field, $A_1$ is the amplitude of the modulation, and $\hat d$ its direction. It is easy to check that such modification affects not only the $\ell=1$ angular multipole, but actually {\it all} multipoles equally, and for this reason it is known as a scale independent dipolar modulation. Its main effect is to create {\it correlations} between multipoles $\ell$ and $\ell+1$. Such correlations, as mentioned above, violate isotropy (see Appendix B of \cite{Agullo:2020cvg} for further details).

The Planck team has carried out a likelihood analysis of such a modulation of the CMB, and arrived at constraints on the amplitude and direction of the dipolar modulation in different  bins of multipoles $\ell$.  Surprisingly, the analysis has revealed a non-zero amplitude of the dipolar modulation {\it only} for low multipoles, in the  bin $\left[ 2\,, \, 64 \right]$. The amplitude reported in this bin is $A_1 \approx 0.07$ \cite{ Ade:2015hxq}, and the significance of the detection is greater than 3-$\sigma$. This reveals, not only a significant deviation of the $\Lambda$CDM model, but also that the dipolar modulation is scale-{\it dependent}, since it only appears for low multipoles. Therefore, the simple model (\ref{eqn:gordon}) is insufficient to account for the observed modulation. Finding a mechanism to generate a scale-dependent dipolar modulation, without introducing other undesired effects, has challenged the imagination of theorists during the last decade \cite{Dai:2013kfa}.

\par
\begin{figure}
    \centering
    \includegraphics[width=0.8\textwidth]{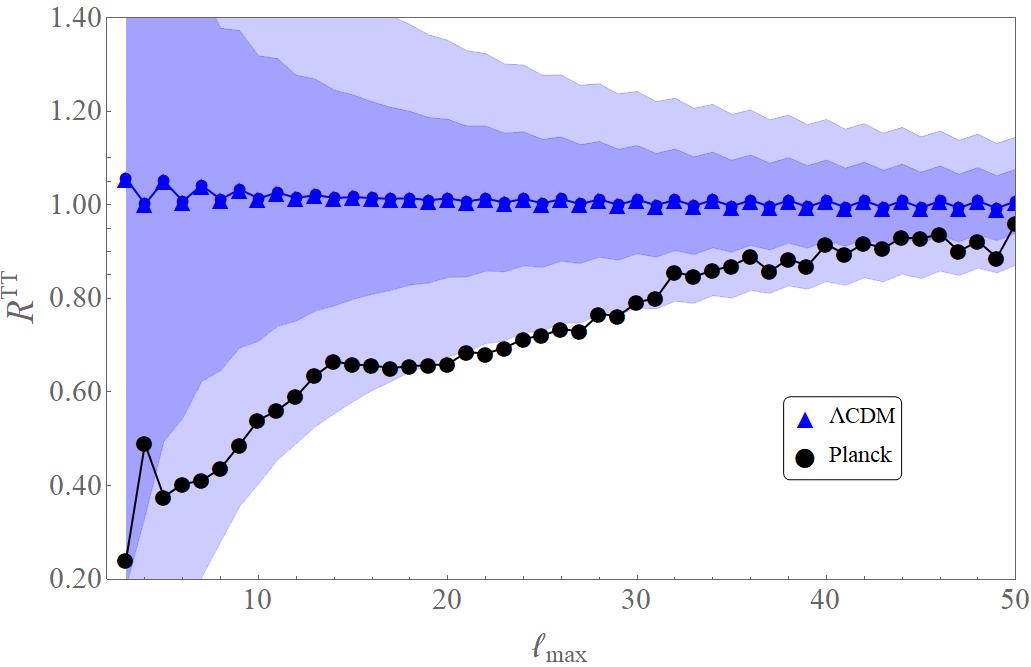}
    \caption{$R^{TT}(\ell_{max})$ generated in standard model (blue) along with 2-$\sigma$ shaded contours arising from cosmic variance. Black points are the observations by Planck. The observed value of $R^{TT}(\ell_{\rm max})$ for most points is lower than the predictions of the standard model by more than 1-$\sigma$.}
    \label{fig:parity}
\end{figure}

\subsection{Parity anomaly}

Observations from both WMAP and Planck have  found a preference for odd parity two-point correlations, as opposed to the predictions of the  standard $\Lambda$CDM model, which predicts that the primordial perturbations generated in our universe are parity neutral. 
The parity of the primordial perturbations can be studied by analyzing the multipoles in the range $[2\, , \, 50]$, known as the Sachs-Wolfe plateau. This  range of multipoles corresponds to long wavelength perturbations which entered the horizon in the recent past, and hence have been relatively unmodified  by late time physics. 
The asymmetry in the parity can be quantified using the estimator 
\begin{equation}\label{eqn:RTT}
    R^{TT}(\ell_{\rm max})\, =\, \frac{D_+(\ell_{\rm max})}{D_-(\ell_{{\rm max}})}\, ,
\end{equation}
where $D_{\pm}(\ell_{\rm max})$ quantifies the sum of power contained in even ($+$) or odd ($-$) multipoles,  up to $\ell_{max}$. More specifically, $D_{\pm}(\ell_{\rm max})$ are defined as
\begin{equation}
    D_{\pm}(\ell_{\rm max})\, =\, \frac{1}{\ell_{tot}^{\pm}}\, \sum_{2,\ell_{max}}^{\pm}\, \frac{\ell (\ell\,+\,1)}{2\,\pi}\, C_\ell
\end{equation}
where the $+$ or $-$ signs on the right refer to the fact that we include  only  even or odd multipoles in the sum, respectively, and $\ell_{tot}^{\pm}$ refers to the total number of multipoles in the sum. 
Fig. \ref{fig:parity} illustrates that CMB data in the multipole range of $[2\, , \, 50]$ shows a clear preference for odd parity compared to the parity neutral, {\it i.e.~}$R^{TT}(\ell_{max}) =\, 1$, prediction of standard model. Although this anomaly, as well as the anomaly in the lensing amplitude discussed in the next subsection, are not as severe as the previous ones due to their lower statistical significance ($\lesssim 2$-$\sigma$), we will later argue that they may be related to the power suppression. 

\par
\subsection {Lensing amplitude anomaly}
The cosmic microwave background radiation undergoes lensing by the intervening distribution of matter, as it propagates from the surface of last scattering to us. An important observable in the CMB, in addition to temperature and polarization, is the lensing potential. From the CMB maps, Planck has reconstructed the lensing potential and computed its power spectrum \cite{Aghanim:2018oex}. The effect of lensing is the smoothing of CMB power spectrum at small angular scales. The amount of smoothing observed in the CMB angular power spectrum should be consistent with the smoothing derived from the power spectrum of the lensing potential. 
In order to check this consistency, \cite{Aghanim:2018eyx} introduced a test-parameter, known as the lensing parameter $A_L$, that multiplies the lensing power spectrum. Theoretically, the value of lensing parameter should be $A_L\, =\, 1$, and in fact 
Planck assumes this value during the process of parameter estimation. However, if $A_L$ is left as a free parameter,   along with the six parameters of the $\Lambda$CDM model, in the Markov Chain Monte Carlo (MCMC) analysis,  one finds that $A_L\, =\, 1.243\, \pm\, 0.096$ for  $Planck TT + lowE$ data,  which is more than 2-$\sigma$ away from one. If the reconstructed lensing data is also used, along with Planck $EE$ and $TE$ data, then the lensing parameter is consistent with 1 within 2-$\sigma$. 
\par

A key feature of the anomalies discussed above, except perhaps for lensing anomaly, is that they appear clearly associated with  the largest angular scales we can observe. This suggests a common origin in  primordial  physics for these diverse set of anomalies. The next section introduces a proposal for a mechanism that can provide such  common origin, namely  the phenomenon of non-Gaussian modulation. Together with the scale dependence introduced by the quantum bounce of LQC, this mechanism constitutes  a promising  candidate  for the origin of the  anomalies we have just described. 


\section{Non-Gaussian modulation}\label{sec:3}
Temperature anisotropies in the CMB are a consequence of the evolution of photons and other constituents of the universe in a perturbed spacetime. Since the observed anisotropies are  small, $\delta T/\bar T \sim 10^{-5}$,  perturbation theory is an appropriate tool.\footnote{Strong non-linearities are important at late times in the universe during structure formation, but not to explain the CMB.}
If the primordial perturbations in the metric generated in the early universe were exactly linear, then only those perturbations with wavelengths smaller than the radius of the Hubble  horizon today would be able to affect the CMB. On the contrary, non-linear effects, generically known as non-Gaussianty, couple modes of different wavelengths, and make it possible that  primordial perturbation with wavelengths larger than the Hubble radius today can impact what we observe in the CMB 
\cite{Schmidt:2010gw, Schmidt:2012ky, Jeong:2012df,Dai:2013kfa,Agullo:2015aba,Adhikari:2015yya}. We will refer to this phenomenon as non-Gaussian modulation of the CMB. 
Since  long wavelength, super-horizon modes do not evolve with time, we could treat them as  spectator modes, whose role is to influence, or bias, sub-horizon modes. 
\par
Primordial perturbations are random variables with zero mean and a variance characterized by the two-point correlations discussed in the previous section. We will show that one consequence of the coupling between super-horizon and sub-horizon wavelengths is to modify the two-point correlation functions \cite{ Schmidt:2012ky,Agullo:2015aba,Adhikari:2015yya}. Though the mean value of the primordial perturbations is not modified, the variance is, in such a way that certain features in the CMB are more likely to be observed than in the absence of non-Gaussian correlations, and consequently they should not be considered as anomalous. 
In this section, we will describe the essential features of the mechanism of non-Gaussian modulation. We will split the discussion in  two parts: in the first one, we will discuss the modulation of the primordial power spectrum due to non-Gaussian correlations with a spectator mode,  and in the second part, we describe the effect of such a modulation on the CMB  $TT$ angular power spectrum.
\subsection{Non-Gaussian modulation of primordial perturbations}
We are interested in computing the two-point correlation function of the curvature perturbation ${\mathcal R}_{\vec k}(\eta)$ for a mode $\vec k$ that is observable in the CMB, in the presence of a longer wavelength mode ${\mathcal R}_{\vec q}$, when both modes are correlated.  
\par
A convenient and general way to model the effects of non-Gaussian correlations, is to write the curvature perturbations ${\mathcal R}_{\vec k}(\eta)$ in terms of a Gaussian field ${\mathcal R}^G$ as follows \cite{Schmidt:2010gw}
 \be \label{eqn:Rk} {\mathcal R}_{\vec k}(\eta)={\mathcal R}^G_{\vec k}(\eta) +\frac{1}{2}\, \int \frac{d^3q}{(2\pi)^3}\,   f_{NL}(\vec q,\vec k-\vec q)\, {\mathcal R}^G_{\vec q}(\eta)\, {\mathcal R}^G
 _{\vec k-\vec q}(\eta)\, . \ee
The convolution in the integral is the Fourier transform of a quadratic  combination of   ${\mathcal R}^G$ in position space, and is the source of the non-Gaussian character of ${\mathcal R}_{\vec k}(\eta)$, and the function $f_{NL}(\vec k_1,\, \vec k_2)$ contains the information about the strength and details of the non-Gaussianty. The goal of this equation is simply to parameterize the non-Gaussianty in a simple and tractable way, while the form of the function $f_{NL}(\vec k_1,\, \vec k_2)$ is expected to come from a concrete microscopic model of the early universe.

Statistical isotropy and homogeneity implies that the function $f_{\rm NL}(\vec k_1,\, \vec k_2)$ depends only on the modulus of  the two wavenumbers involved, $k_1\equiv |\vec k_1|$ and $k_2\equiv |\vec k_2|$, and on the (cosine of) angle between them, $\mu$: $f_{\rm NL}(\vec k_1,\, \vec k_2)=f_{\rm NL}(k_1,\, k_2,\mu)$. From it, the three-point correlation function is given by $\langle  {\mathcal R}_{\vec k_1}{\mathcal R}_{\vec k_2} {\mathcal R}_{\vec k_3} \rangle=(2\pi)^3\, \delta(\vec k_1+\vec k_2+\vec k_3)\, B_{\mathcal R}(\vec k_1,\vec k_2)$, where the bispectrum $B_{\mathcal R}(\vec k_1,\vec k_2)$ is 
\be \label{bisp} B_{\mathcal R}(\vec k_1,\vec k_2)= f_{NL}(\vec k_1,\vec k_2) \, [P_{\mathcal R}(\vec k_1)P_{\mathcal R}(\vec k_2)+1
\leftrightarrow 3+2
\leftrightarrow 3]\, ,\ee
and $P_{\mathcal R}(\vec k$) is the power spectrum of ${\mathcal R}^G$, defined as 
\be \label{ps} \langle {\mathcal R}^G(\vec k_1){\mathcal R}^{G^\star}(\vec k_2) \rangle =(2\pi)^3\, \delta(\vec k_1-\vec k_2)\, P_{\mathcal R}(\vec k_1)\, .\ee
The dimensionless power spectrum is defined as ${\mathcal P}_{\mathcal R}(\vec k)\, =\, k^3\,P_{\mathcal R}(\vec k)/2\pi^2$.
\par
Our goal is to compute the two-point function of ${\mathcal R}_{\vec k}$ in presence of the spectator mode ${\mathcal R}_{\vec q}$. Using (\ref{eqn:Rk}), one obtains 
\bea \label{modtp} \langle {\mathcal R}_{\vec k_1}{\mathcal R}^{\star}_{\vec k_2}\rangle |_{{\mathcal R}_{\vec q}}= \langle {\mathcal R}^G_{\vec k_1}{{\mathcal R}^G}^{\star}_{\vec k_2}\rangle &+& \frac{1}{2}\, \int \frac{d^3q'}{(2\pi)^3}\, f_{NL}(\vec{q'} ,\vec k_1-\vec{q'})\, \langle{\mathcal R}^G_{\vec{q'}}\, {\mathcal R}^G_{\vec k_1-\vec{q'}}\, {{\mathcal R}^G}^{\star}_{\vec k_2}\rangle \nonumber \\ &+& \frac{1}{2}\, \int \frac{d^3q'}{(2\pi)^3}\, f_{NL}(\vec{q'},\vec k_2-\vec{q'})\, \langle{\mathcal R}^G_{\vec k_1}\, {{\mathcal R}^G}^{\star}_{\vec{q'}}\, {{\mathcal R}^G}^{\star}_{\vec k_2-\vec{q'}}\rangle+\mathcal{O}(f_{NL}^2).\,
  \eea
In order to evaluate the impact of the spectator modes ${\mathcal R}^G_{\vec q}$, it must be taken out of the statistical average
  \bea \label{ngmod} \langle {\mathcal R}_{\vec k_1}{\mathcal R}^{\star}_{\vec k_2}\rangle |_{{\mathcal R}_{\vec q}}&=& (2\pi)^3\, \delta(\vec k_1-\vec k_2)\, P_{{\mathcal R}}(\vec k_1)\nonumber \\&+&   f_{NL}(\vec k_1, - \vec k_2)\, \frac{1}{2}\, \big(P_{{\mathcal R}}(\vec k_1)+P_{{\mathcal R}}(\vec k_2)\big)\, {\mathcal R}_{\vec q}+\cdots \, .\eea
where the trailing dots indicate terms that are higher order in non-Gaussianity, and will be subdominant. 
\par
It is interesting to note the following facts about the above expression. First of all, non-Gaussianity leads to a modulation of the primordial power spectrum, and the strength of modulation depends on both the size and shape of $f_{NL}(\vec k_1,\,\vec k_2)$, as well  as the size of the spectator mode ${\mathcal R}_{\vec q}$. Secondly, statistical isotropy and homogeneity constrain the wavenumber of the spectator mode to be $\vec q\, =\, \vec k_1\, -\, \vec k_2$. In other words, this is the only mode that can affect the two-point correlation function between $\vec k_1$ and $\vec k_2$. Additionally, the effect of the modulation is to introduce ``non-diagonal'' elements in the two-point function, i.e., terms not proportional to $\delta(\vec k_1-\vec k_2)$. But recall that such non-diagonal terms break homogeneity and isotropy. It is not surprising that we see deviations from these fundamental symmetries, since we are not averaging over the spectator mode: such average would make those terms disappear, since $\langle {\mathcal R}^G_{\vec q}\rangle\, =\, 0$. 
But, as it happens for the magnitude of the temperature anisotropies, the quantity that is more interesting for observations is the typical value of such term, and not only its statistical average. 

\subsection{Non-Gaussian modulation of CMB}

The primordial perturbations  ${\mathcal R}_{\vec k}$ are related to the CMB multipole coefficients $a_{\ell m}$ through the relation 
 \be \label{alm}  a_{\ell m}=4\pi \int \frac{d^3k}{(2\pi)^3}\, (-i)^{\ell}\, \Delta_{\ell}(k)\, Y^*_{\ell m}(\hat k)\, {\mathcal R}_{\vec k} \, , \ee 
where  $ \Delta_{\ell}(k)$ are the CMB temperature transfer functions, which encode the complications of the post-inflationary evolution of the perturbations. From this equation, one can compute the  covariance matrix 
\be \label{aaa} \langle a_{\ell m} a^{\star}_{\ell' m'}\rangle= (4\pi)^2\int \frac{d^3k_1}{(2\pi)^3}\int \frac{d^3k_2}{(2\pi)^3}\, (-i)^{\ell-\ell'}\, \Delta_{\ell}(k_1)\, \Delta_{\ell'}(k_2)\, Y^*_{\ell m}(\hat k_1)\, Y_{\ell' m'}(\hat k_2)\, \langle {\mathcal R}_{\vec k_1}{\mathcal R}^{\star}_{\vec k_2}\rangle |_{{\mathcal R}_{\vec q}}\, ,\ee
which is obtained from the  two-point functions of the curvature perturbations given in (\ref{ngmod}). 
Upon  expanding $f_{NL}$ in terms of Legendre polynomials, $f_{NL}(k_1,q,\mu)=\sum_L G_L(k_1,q)\, \frac{2L+1}{2}\, P_L(\mu)$, and using the multipole expansion ${\mathcal R}^G_{\vec q}=\sum_{L'M'} {\mathcal R}^G_{L'M'}(q)\, Y_{L'M'}(\hat q)$, one can  write (\ref{aaa}) as \cite{Agullo:2020cvg}
 \be \label{abiposh} \langle a_{\ell m} a^*_{\ell' m'}\rangle= C_{\ell}\, \delta_{\ell\ell'}\delta_{m m'} +(-1)^{m'}\,\sum_{LM} A^{LM}_{\ell\ell'}\,  C^{LM}_{\ell m \ell' -m'}\,. \ee
 The above expression consists of two terms. 
 The first term is the usual temperature power spectrum that is  diagonal in $\ell$ and $m$. 
 The second term arises from the non-Gaussian modulation and, as before, introduces non-diagonal terms.  $C^{LM}_{\ell m \ell' m'}$ are  Clebsch-Gordan coefficients, and the  information about the primordial non-Gaussianty is encoded in the coefficients
\bea  \label{valueBipoSH} A^{LM}_{\ell\ell'} &=&\frac{4}{(2\pi)^3}\,  \int dk_1\, k_1^2\,{dq\, q^2} \, (-i)^{\ell-\ell'}\,\Delta_{\ell}(k_1)\,  \Delta_{\ell'}(k_1)\, P_{\mathcal R}(k_1) \, \,  G_L(k_1,q)\, {\mathcal R}^G_{LM}(q)\, \nonumber \\ &\times& C^{L0}_{\ell 0 \ell' 0}\,  \sqrt{\frac{(2\ell+1)(2\ell'+1)}{4\pi\, (2L+1)}} \, .\eea
These coefficients  are known as bipolar spherical harmonic (BipoSH) coefficients \cite{Hajian_2003, Joshi:2009mj}. As we shall see, the BipoSH coefficients provide a convenient way to  organize the effects of the non-Gaussian modulation. 


\par
 
The Clebsch-Gordon coefficients present in the above expressions enforce certain properties on the BipoSH coefficients. In particular, Clebsch-Gordon coefficients $C^{LM}_{\ell_1,m_1,\ell_2,m_2}$ are nonzero only if $\ell_1\,+\,\ell_2\,\geq\, L\,\geq\,|\ell_1\,-\,\ell_2|$ and if $M\,=\, m_1\,+\,m_2$. This, together with properties of the Clebsch-Gordon coefficient $C^{L0}_{\ell 0 \ell' 0}$, implies that, if
\begin{enumerate}[i.]
    \item\; $L\,=\,0$, then $\ell_1\,=\,\ell_2$
    \item\; $L\,=\,1$, then $|\ell_1\,-\,\ell_2|\,=\, 1$
    \item\; $L\,=\,2$, then $|\ell_1\,-\,\ell_2|\,=\, 0,\,2$, {\rm etc.}
\end{enumerate}
 Thus, a non-zero value of $A^{LM}_{\ell\ell'} $ for $L\,=\,0$ can be absorbed in the diagonal angular power spectrum $C_{\ell}$. A non-zero value of   $A^{LM}_{\ell\ell'} $  for $L=1$ induces correlations between multipoles $\ell$ and $\ell+1$, or in other words, a dipolar modulation. $L=2$ induces a quadrupolar modulation, etc. The presence of a large dipolar or higher multipole modulation would appear in the CMB as correlations between multipoles $\ell$ and $\ell+1$, which implies a  departure from  isotropy, as described in section \ref{sec:2}. This departure from isotropy is a consequence of the concrete realization of the spectator mode $\mathcal{R}_{\vec q}$ in our local universe. One would need to average among the observation of the CMB from distant places in the cosmos to conclude that such violation of isotropy is not fundamental, but rather the imprint of strong correlations with super-horizon modes $\mathcal{R}_{\vec q}$.

 \par
Two remarks are  in order now. 
\begin{enumerate}[i.]
\item The strength of non-Gaussian modulation is dictated by the size of $f_{NL}(k_1,\,q,\,\mu)$. But it is the dependence of $f_{NL}$ on $\mu$, the cosine of the angle between $\vec k_1$ and $\vec q$, what determines the relative size of the BipoSH coefficients for different $L$'s, i.e., the ``shape'' of the modulation. On the other hand, the dependence of $f_{NL}$ on the moduli  $k_1$ and $q$ determines the $\ell$-dependence of the modulation. The two multipoles should not be confused: the $L$-dependence dictates the shape of the modulation, while the $\ell$-dependence controls the variation of the amplitude of the modulation at different angular scales in the CMB. 
The non-Gaussianity generated in slow-roll inflation is small and nearly scale invariant. Hence, the strength of modulation generated is also quite small. Since the anomalies observed in the CMB are scale-dependent, we need a scenario with a strongly scale-dependent and large non-Gaussianity. Such scale-dependence is also needed to explain why we have not observed non-Gaussian correlations directly in the CMB, since a strong  scale-dependence can make these correlations large only when at least one super-horizon mode is involved. In such situation, we could only observe the indirect effects that the non-Gaussian correlations induce in the  CMB.

\item $A^{LM}_{\ell \ell'}$ given in (\ref{valueBipoSH}) depend on the mode ${\mathcal R}^G_{\vec q}$. Since, ${\mathcal R}^G_{\vec q}$ is a random variable, we cannot predict the exact value of $A^{LM}_{\ell \ell'}$. We can only compute the standard deviation of the BipoSH coefficient, {\it i.e~} 
\be \label{varalpha}    \sqrt{\langle|A^{L M}_{\ell \ell' }|^2\rangle}= \left[\frac{1}{2\pi} \, \int dq\, q^2 \, P_{{\mathcal R}}(q) \, |\mathcal{C}_{\ell \ell'}^L(q)|^2\, \right]^{1/2} \times C^{L0}_{\ell 0 \ell' 0}\,  \sqrt{\frac{(2\ell+1)(2\ell'+1)}{4\pi\, (2L+1)}}\, , \ee
where 
 \be \label{Ctilde}  \mathcal{C}_{\ell \ell'}^{L}(q)\equiv \frac{2}{\pi}\int dk_1 \, k_1^2 \, (i)^{\ell-\ell'} \, \Delta_{\ell}(k_1)\Delta_{\ell'}(k_1) \, P_{{\mathcal R}}(k_1)\, G_L(k_1,q) \, .\ee
These are the typical values that the BipoSH coefficients are expected to take in the sky. If these values are large, the effects they entail should be expected in the CMB or, more precisely, they would have a large $p$-value and should not be considered anomalous. 

\end{enumerate}
 
\section{Loop quantum cosmology}\label{sec:4}
LQC describes the spacetime geometry in the quantum language of loop quantum gravity. The quantum geometry of the early universe is sourced by scalar field, which is responsible for driving the universe  to an inflationary phase after the bounce. The bounce introduces a new physical scale in the problem, which can be defined either from the value of the energy density or from the Ricci scalar at the bounce. Perturbations, both scalar and tensor,  are sensitive to this new scale, and their propagation across the bounce amplifies them, for the same reason that propagation across the inflationary phase  does. As a consequence, perturbations reach the onset of inflation in an excited and  non-Gaussian state, rather than the Bunch-Davies vacuum commonly postulated. These excitations lead to a strongly scale dependent and enhanced  power spectrum and bispectrum of primordial perturbations.  In this section, we will briefly review some of the essential features of perturbations generated in LQC, and in  next section we will describe how these features  can account for the anomalous signals observed in the CMB. For further details, see \cite{Agullo:2012sh,Agullo:2012fc,Agullo:2013ai,Agullo:2015tca,Agullo:2017eyh}

\subsection{Background dynamics and free evolution of perturbations}
\begin{figure}
    \centering
    \begin{tabular}{cc}
    \includegraphics[width=0.46\textwidth]{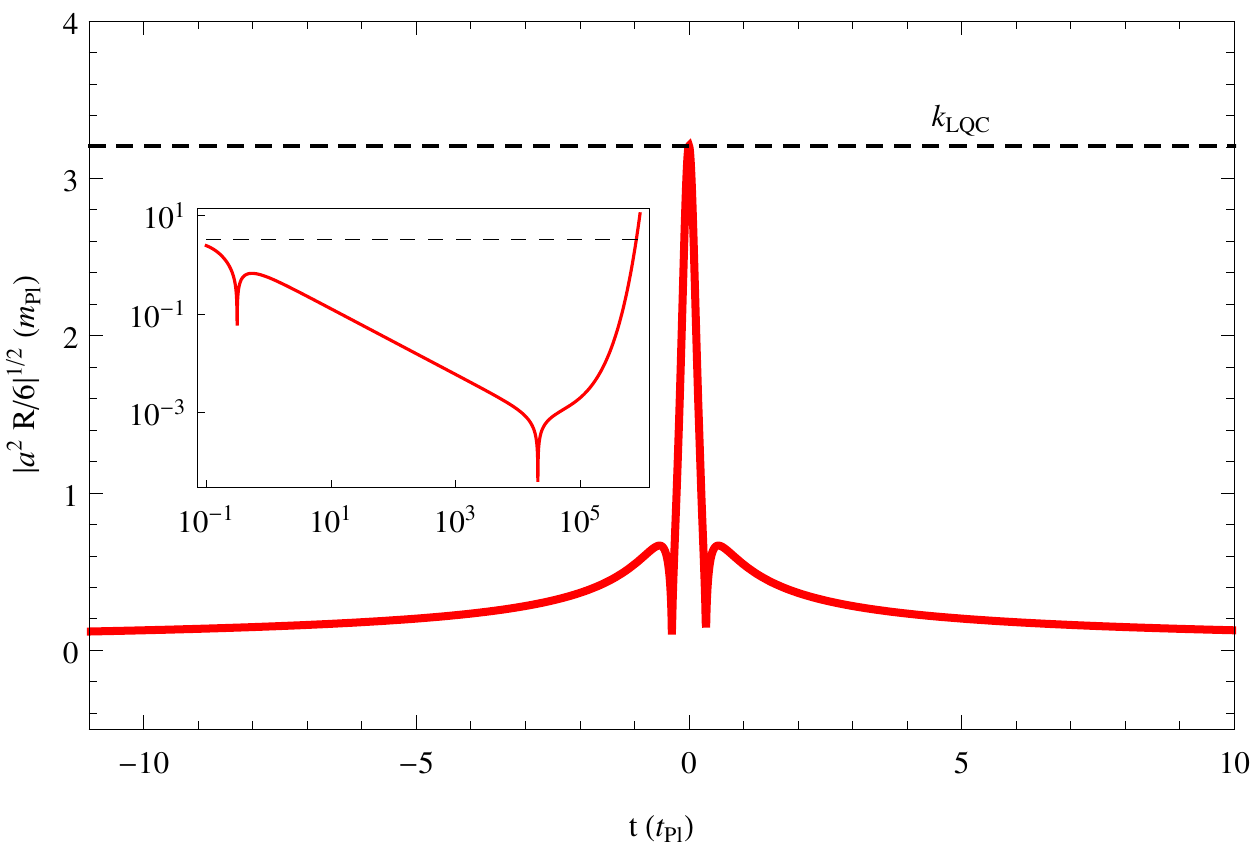} &
    \includegraphics[width=0.46\textwidth]{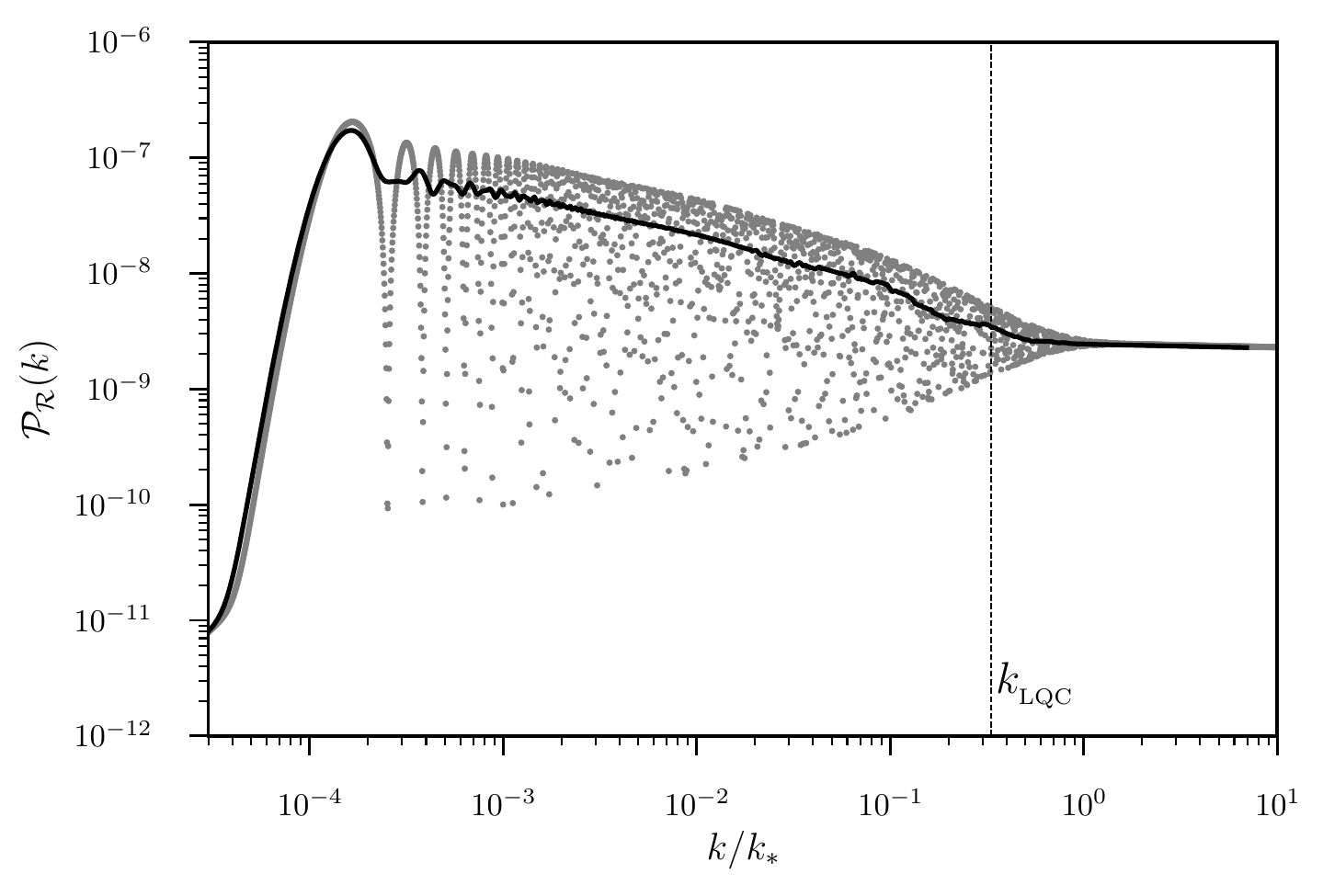}
    \end{tabular}
    \caption{Evolution of relevant scales in LQC (left). Primordial power spectrum generated in LQC (right). }
    \label{fig:lqc}
\end{figure}
Consider a  spatially flat Friedmann-Lemaitre-Robertson-Walker spacetime. 
We shall describe the perturbations following the dressed metric approach. This approach has been discussed in  \cite{Agullo:2012sh,Agullo:2012fc,Agullo:2013ai,Agullo:2015tca,Agullo:2017eyh}  (for a recent review, see, \cite{Agullo:2016tjh}) and  we refer the reader to these references for details omitted here. For the purpose of this article, it suffices to say that we consider perturbations as test fields propagating on the background described by the effective equations of LQC. The essential features of perturbations generated in LQC, can be summarized using the Fig. \ref{fig:lqc}. This figure plots $a\sqrt{|R/6|}$ as a function of time, where $a$ refers to the scale factor and $R$ is the Ricci scalar. In making this plot we have worked with a scalar field governed by a quadratic potential minimally coupled to gravity. Similar results are obtained for other potentials \cite{Bonga:2015kaa,Bonga:2015xna,Zhu:2017jew}. The peak of the curvature occurs at the bounce. This sets a characteristic scale in LQC denoted by $k_{LQC}\, \equiv\,a(t_B)\,\sqrt{R(t_B)/6}\approx a(t_B)\,\sqrt{\kappa\,\rho_B} $, where $t_B$ indicates the time of bounce and $\rho_B$ is the energy density of the scalar field at the time of the bounce.  As the inset in the plot shows, inflation sets in at late time. Perturbations are in an adiabatic regime before the bounce, and we consider they start in an adiabatic vacuum at those early times (see e.g.~\cite{Agullo:2015tca,Agullo:2014ica,deBlas:2016puz,Ashtekar:2016wpi,ElizagaNavascues:2019itm,ElizagaNavascues:2020fai,Martin-Benito:2021szh} for other choices of initial state). As the perturbations evolve across the bounce, modes with wavenumbers  $k\,\lesssim\,k_{LQC}$ are excited.  
These excitations get further amplified as they cross the curvature scale during inflation. Wavenumbers that are ultraviolet compared to $k_{LQC}$, $k\,>\,k_{LQC}$,  are not excited during the bounce, and remain in the adiabatic vacuum at the onset of inflation. Hence, only for those modes one recovers the familiar Bunch-Davies vacuum at the onset of inflation, while more infrared modes keep memory of the bounce.  Consequently, as shown in Fig. \ref{fig:lqc}, the power spectrum of curvature perturbations shows a strong scale dependence at infrared scales, while approaches the more familiar scale invariant shape for large $k$'s. In particular, we see that the power spectrum for infrared modes $k\,\lesssim\,k_{LQC}$ is enhanced and oscillatory. In the extreme infrared limit modes are neither excited during bounce nor during inflation, and this leads to a power spectrum which scales as $k^2$. The  scale at which these effects appear in the CMB depends on the physical size of the mode $k_{LQC}$ today, compared to the Hubble scale  (recall that  the physical wavenumber scales with  time as $k_{LQC}/a(t)$). This depends on the expansion accumulated---i.e., the number of $e$-folds $N$---from the time of the bounce until the end of inflation. This is a free parameter in LQC. In this article, we investigate whether there is a value of $N$ for which this model can explain the origin of the anomalies in the CMB.

\subsection{Generation of primordial non-Gaussianity}
The dressed metric approach was extended beyond linear perturbation theory in  \cite{Agullo:2017eyh}, and we provide here a short summary.  Primordial curvature perturbations whose wavenumbers are comparable to or smaller than $k_{LQC}$ not only get excited, as described above, but also become non-Gaussian as they cross the bounce. The non-Gaussianity thus generated is further enhanced as the perturbations cross the horizon during inflation. Equal-time three-point functions are computed using time dependent perturbation theory, generalizing the pioneering  calculations in \cite{Maldacena:2002vr} to bouncing geometries:
\begin{equation}\label{eqn:inin}
    \langle 0| \hat{\mathcal R}_{\vec k_1}(\eta_e)\,\hat{\mathcal R}_{\vec k_2}(\eta_e)\,\hat{\mathcal R}_{\vec k_3}(\eta_e) |0\rangle\, =\, i\,\int_{\eta_i}^{\eta_e}\, {\rm d}\eta'\,  \langle 0| \, \biggl[ \hat{\mathcal R}_{\vec k_1}(\eta)\, \hat{\mathcal R}_{\vec k_2}(\eta)\, \hat{\mathcal R}_{\vec k_3}(\eta),\,
\hat{\mathcal H}_{int}(\eta')\biggr]
| 0\rangle,
\end{equation}
where $\hat{\mathcal H}_{int}$ is the interaction Hamiltonian (whose lengthy expression can be found in \cite{Agullo:2017eyh}), $\eta_i$ refers to the time at which initial conditions are imposed and $\eta_e$ is the time at which the correlation is evaluated. Usually, $\eta_e$ is chosen at the end of inflation, after all the three modes have crossed the Hubble radius. With the knowledge of the background dynamics and the initial conditions, we can exactly evaluate the three-point function and hence obtain the function $f_{NL}(\vec k_1,\, \vec k_2)$ which characterizes the non-Gaussianity. Our exact computations reveal that the non-Gaussianity generated in LQC is strongly scale dependent, large and  oscillatory, similar to the power spectrum. As for the power spectrum, the non-Gaussianity quickly approaches the inflationary result for wave numbers $k\,>\,k_{LQC}$ \cite{Agullo:2017eyh, Sreenath:2019uuo}, and in particular they become negligibly small when the moduli of the three wave numbers $\vec k_1$, $\vec k_2$ and $\vec k_1-\vec k_2$ are larger than $k_{LQC}$, in such a way that they are too small to be observed directly in the CMB. However, the non-Gaussianity becomes large when at least one of the modes involved are infrared, $k<k_{LQC}$, or equivalently, when one of the modes has wavelength larger than the Hubble radius today.  These are the correlations which can account for the CMB anomalies, as we argue in the next section.  
\par
The strong oscillatory character of non-Gaussianity generated in LQC, makes it computationally difficult to obtain an exact evaluation  of Eqn.~(\ref{varalpha}). For this reason, in this work, rather than working with the exact numerically-evaluated non-Gaussianity, we shall work with an analytical approximation  derived in  \cite{Agullo:2017eyh} 
\begin{equation}\label{eqn:shape}
    f_{NL}(k_1,\, k_2,\, k_3)\,\simeq \mathfrak{f}_{NL}\, e^{-\alpha\,(k_1\,+\, k_2\,+\,k_3)/k_{LQC}}, 
\end{equation}
where $\alpha\, =\, 0.647$ and $\mathfrak{f}_{NL}\approx 2750$, and $k_3=k_1\sqrt{1+\frac{k_2^2}{k_1^2}+2\mu \frac{k_2}{k_1}}$. The value of  $\alpha$ is determined from the behavior of the scale factor around the time of the bounce, while the amplitude $\mathfrak{f}_{NL}$ is determined from numerical simulations \cite{Agullo:2017eyh}.   As showed in \cite{Agullo:2017eyh}, this expression  provides a good approximation for the non-Gaussianity generated in LQC, and is significantly easier to manipulate. This approximation,  however, neglects the  oscillatory nature of  $f_{NL}(k_1,k_2,\mu)$ with $k_1$ and $k_2$. The oscillations will generically reduce the size of the effects we describe below. Therefore, the numbers obtained in the next section should be understood as an upper bound for the predictions of LQC, rather than an exact result. This is the main technical limitation of our analysis, and it arises from the highly oscillatory nature of the perturbations.

\section{Results}\label{sec:5}
In this section, we shall put the previous results together and compute the root mean square value of the BipoSH coefficients  generated in LQC from Eqn.~(\ref{varalpha}). We will show that the BipoSH coefficients generated in this model are non-zero and have the appropriate magnitude and scale dependence as demanded by observations.

\subsection{Monopolar modulation -- power suppression}\label{sec:monopole}

\begin{figure}
    \centering
    \includegraphics[width=0.8\textwidth]{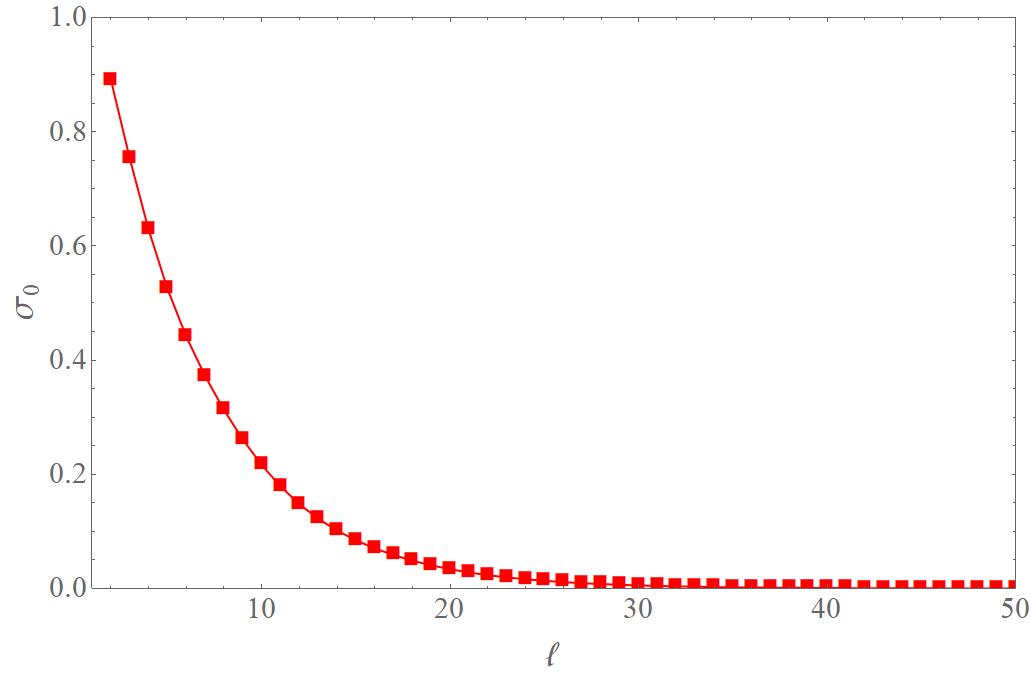}
    \caption{Root-mean-square of the monopolar modulation $\sigma_0(\ell)$ generated in LQC. Note the dependence of $\sigma_0$ on $\ell$. The scale dependence introduced by the bounce in LQC makes the effects of the modulation significant only for $\ell \lesssim 30$.}
    \label{fig:sigma0}
\end{figure}

We first consider the monopolar term ($L\,=0$). The properties of the Clebsch-Gordon coefficients for $L=0$  impose the constraints $\ell\, =\, \ell'$ and $m\,=\,-m'$. Therefore, the monopolar modulation  introduces an isotropic shift in the value of $C_\ell$, although the shift can be different for different values of $\ell$. More concretely,  the modulated power spectrum $C^{mod}_\ell$ is  given by
\begin{equation} \label{modC}
    C^{mod}_\ell\, =\, C_\ell\,\biggl(\, 1\, -\, \frac{(-1)^{\ell}}{C_\ell}\,\frac{A^{00}_{\ell\,\ell}}{\sqrt{2\,\ell\,+\, 1}}\,\biggr). 
\end{equation}
Note that $A^{00}_{\ell\,\ell}$ can be either positive or negative, leading to an enhancement or suppression of $C^{mod}_{\ell}$ with respect to $C_\ell$. As explained before, we cannot predict the exact value of $A^{00}_{\ell\,\ell}$. The interesting quantity is rather the root-mean-square value of the modulation: 
\begin{equation}
    \sigma_0^2(\ell)\, =\, \frac{1}{C_\ell^2}\,\frac{ \langle|A^{00}_{\ell\,\ell}|^2\rangle}{2\,\ell\, +\, 1}\, =\, \frac{1}{C^2_{\ell}}\, \frac{1}{8\pi^2}\,  \int dq\, q^2 \, P_{\mathcal R}(q) \, |\mathcal{C}_{\ell \ell}^0(q)|^2\, , 
\end{equation}
where $C^0_{\ell\,\ell}(q)$ was defined in Eqn. (\ref{Ctilde}). This quantity determines the typical size and scale dependence of the monopolar modulation expected in the CMB.  A large value of $\sigma_0$ would make deviations from the unmodulated power spectrum, $C_\ell$ more likely to be observed in the CMB.  
The result of our calculations, using the power spectrum and the form of $f_{NL}(k_1, q,\mu)$ described in the previous section, is plotted in Fig. \ref{fig:sigma0}. 
\par
\begin{figure}
    \centering
    \begin{tabular}{cc}
    \includegraphics[width=0.46\textwidth]{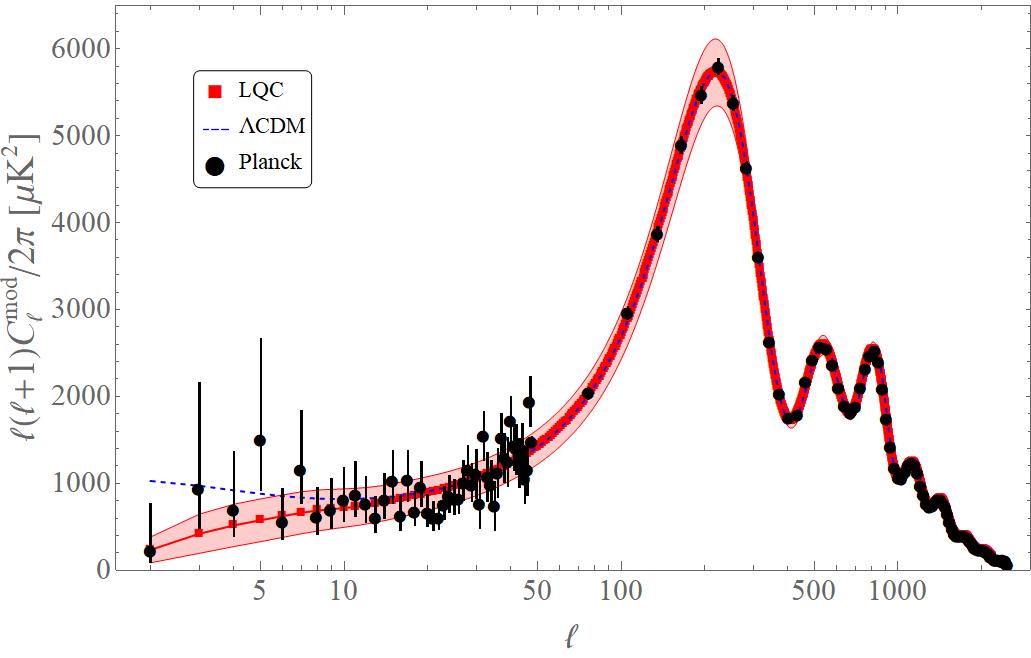}&
    \includegraphics[width=0.46\textwidth]{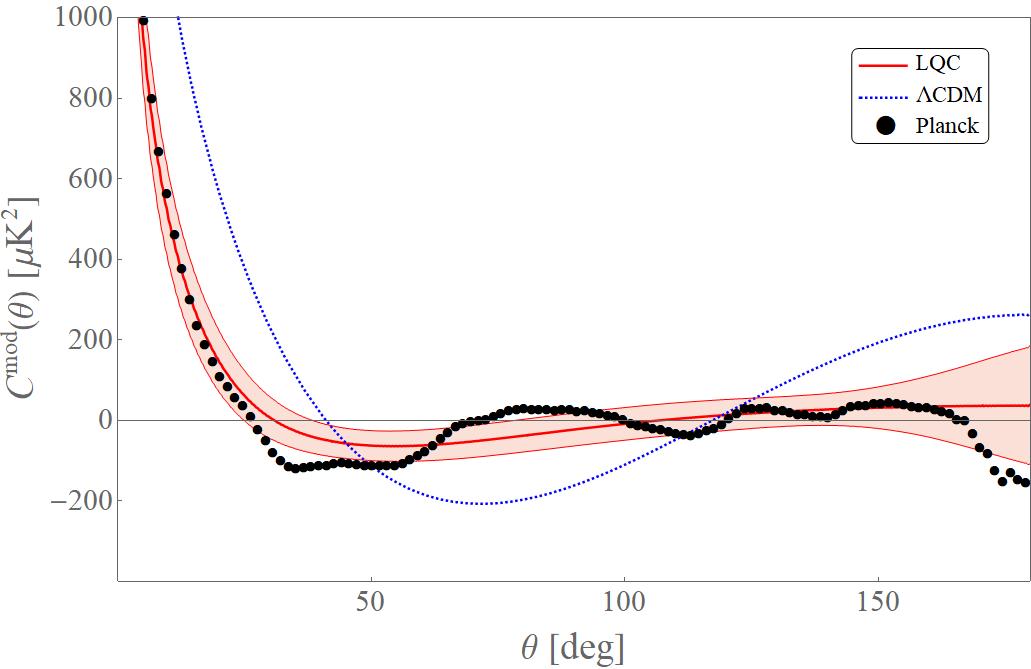}
    \end{tabular}
    \caption{Form of the modulated power spectrum for a typical value of the suppression.  
    The figure shows that the monopolar modulation can account for the suppression of power at multipoles lower than $\ell\, \approx\,30$ (left). In the right panel, we also see that the suppression translates to a very low real space power spectrum $C(\theta)$ for $\theta \geq 60^{\circ}$. The shaded region shows cosmic variance, and the black dots data from Planck.}
    \label{fig:modulated_spectrum}
\end{figure}

We will assume that the probability distribution for the  modulation is well approximated by a Gaussian, and hence completely characterized by $\sigma_0(\ell)$. This is a reasonable approximation, since the deviations are expected to be of second order in non-Gaussianity, and therefore very small. With this probability distribution for the monopolar modulation,  we can now investigate the connection with the  power suppression observed in the CMB. In particular, we want to answer the following question: 
what is the $p$-value for the observed value of $S_{1/2}$? 
We obtain that probability to find $S_{1/2}\leq 1209.2$  once the non-Gaussian modulation is taken into account is approximately $16\%$. This is equivalent to say that the observed suppression is around one standard deviation from the mean. Figure \ref{fig:modulated_spectrum} shows the form of the $T$-$T$ power spectrum for a simulation for which the monopolar modulation produces $S_{1/2}$ in agreement with observation, along with the 1-$\sigma$ confidence contour arising from cosmic variance. For comparison, we provide the corresponding quantities arising from the standard model, as well as data from Planck. 

These results show that, in presence of the LQC bounce occurring before inflation, a power suppression as the one we observe in the CMB should not be considered anomalous. It is important to emphasize the precise sense in which the suppression is explained: not because the theory predicts that we should observe a suppression in the CMB, but rather because the probability of observing such suppression is much larger than in the standard $\Lambda$CDM model with Bunch-Davies initial conditions. In this sense, the resolution of the anomaly has precisely the same character as its origin: probabilistic.

 An important check is to confirm that the non-Gaussian effects are not large enough to jeopardize the validity of the perturbative expansion on which the calculations rest. This question was explored in detail in Ref.~\cite{Agullo:2017eyh}, confirming that in LQC perturbation theory does not break down when non-Gaussianty is included. Regarding the non-Gaussian modulation discussed in this paper, we find that the correction to the unmodulated angular power spectrum is not small, and it is in fact a significant fraction of the final result, particularly for the smallest multipoles. The relative  contribution is, however, smaller than one in all our calculations. In quantitative terms, the relative contribution of the non-Gaussian modulation is of order $ \mathfrak{f}_{_{\rm NL}} \sqrt{\mathcal{P}_{\mathcal{R}}}$, which is smaller than one for $\mathfrak{f}_{_{\rm NL}}\sim 10^3$.  More importantly, higher order corrections introduce  additional powers of the power spectrum $\mathcal{P}_{\mathcal{R}}\ll 1$. So the next-to-leading-order correction to the non-Gaussian modulation is of order  $ \mathfrak{f}_{_{\rm NL}} (\mathcal{P}_{\mathcal{R}})^{3/2}$, which is negligible due to the smallness of $\mathcal{P}_{\mathcal{R}}$. Therefore, our results are robust under the addition of higher perturbative corrections.

\subsection{Dipolar modulation}
Next, we discuss the effects of the $L=1$, dipolar modulation, induced by the BipoSH coefficients, $A^{1M}_{\ell\ell+1}$, and compare the results with those reported by Planck. As discussed in section \ref{dipolar}, the Planck team quantifies the dipolar modulation in terms of a scale-dependent amplitude $A_1(\ell)$ \cite{ Ade:2015hxq}, which can be related with the BipoSH coefficients $A^{1M}_{\ell\ell+1}$ as follows. First, define from $A^{1M}_{\ell\ell+1}$ the multipole coefficients $m_{1M}$ by

\begin{equation}
    \label{aa} A_{\ell\ell+1}^{1M}\equiv m_{1M}\, G^1_{\ell\ell+1},\, \hspace{1cm} {\rm where} \hspace{1cm}  G^1_{\ell\ell+1}\equiv (C_{\ell} +C_{\ell+1}) \sqrt{\frac{(2\ell+1)(2\ell+3)}{4\pi\, 3}}\, C^{10}_{\ell,0,\ell+1,0}\, 
\end{equation} 
is called a form factor. The $m_{1M}(\ell)$ defined above can take three values corresponding to $M\,=\, -1,\,0,\,+1$, and in general they  depend on $\ell$. From them, the  amplitude of the dipolar modulation  is defined as
\begin{equation}
    A_1(\ell)\, \equiv\, \frac{3}{2}\,\sqrt{\frac{1}{3\pi}\biggl(|m_{1\,-1}|^2\,+\, |m_{1\,0}|^2\,+\, |m_{1\,1}|^2\,\biggr) }.
\end{equation}
Hence, from the value of the root-mean-square of $A_{\ell\ell+1}^{1M}$ we can obtain the root-mean-square of $A_1(\ell)$. It is given by the expression 
\begin{equation}
     A_1(\ell)=\frac{3}{2}\frac{1}{\sqrt{\pi}}\frac{1}{C_{\ell}^{ \rm mod} +C_{\ell+1}^{ \rm mod}} \sqrt{\frac{1}{2\pi}\,  \int dq\, q^2 \, P_{\mathcal R}(q) \, |\mathcal{C}_{\ell \ell+1}^1(q)|^2}\, ,
\end{equation}
where we have used the modulated  (i.e.\, suppressed) $C_{\ell}^{ \rm mod}$ since, as emphasized in \cite{ Ade:2015hxq}, the dipole amplitude must be evaluated relative to the observed angular power spectrum. Hence, the fact that the observed $C_{\ell}^{ \rm mod}$  are smaller than the ones predicted by $\Lambda$CDM, increases the amplitude of the observed dipole. In this sense, the power suppression and the dipolar modulation are not completely independent. However, the amplitude of the dipole is ultimately dictated from the angular $\mu$-dependence of the primordial non-Gaussianity $f_{NL}(k_1,k_2,\mu)$.

\begin{figure}
    \centering
    \includegraphics[width=0.8\textwidth]{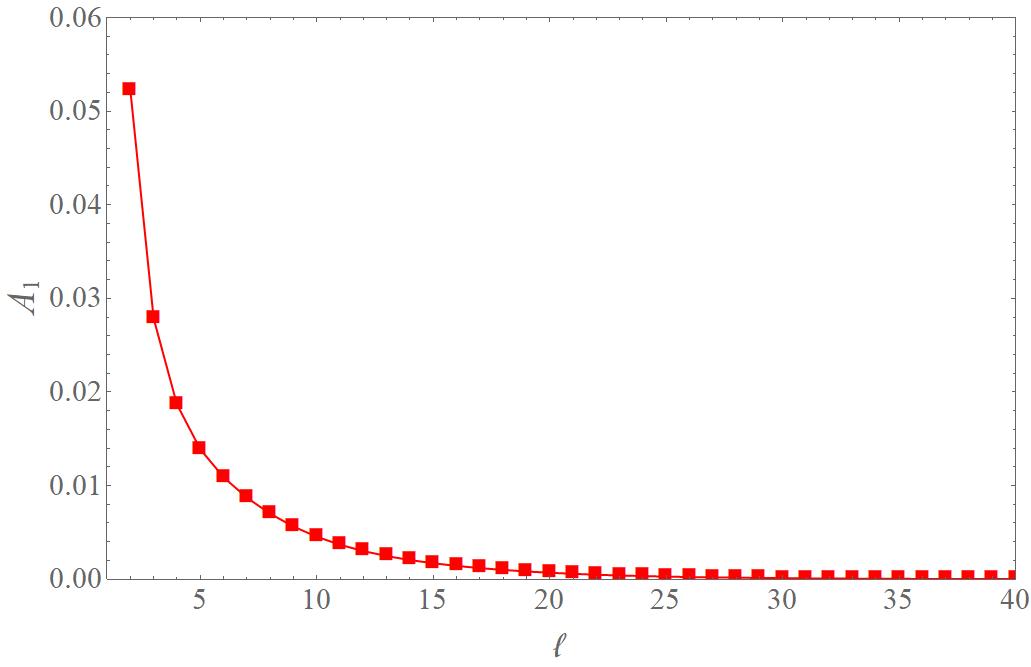}
    \caption{The dipole amplitude $A_1(\ell)$ generated in LQC. Planck reports a value of $A_1\, \approx\, 0.07$ in the multipole bin $[2,\,64]$.}
    \label{fig:dipole}
\end{figure}

The  result for  $A_1(\ell)$ is plotted in Fig.~\ref{fig:dipole}. We find that the dipolar modulation is strongly scale dependent, as a consequence of the scale-dependent nature of the non-Gaussianity. Although Planck observations for $A_1(\ell)$ are limited, in the sense that only its mean value in the range $\ell \in[2,64]$ is reported, the order of magnitude and scale dependence agree with our results.

We have also checked that higher order multipolar modulations, $L=2, 4,\cdots$ have amplitudes significantly smaller than the dipolar one \cite{Agullo:2017eyh}, and therefore additional modulations are not expected in the CMB according to LQC, in agreement with observations. Hence, interestingly, the  form of $f_{NL}(k_1,k_2,\mu)$ derived from LQC produces a hierarchy in the amplitude of the modulations which is dominated by a monopole, and a smaller dipole.

\subsection{Parity and lensing anomalies}

In this subsection we briefly discuss the results for the parity and lensing anomalies. As discussed in previous sections, the statistical evidence for these two features is weaker than the power suppression and the dipolar anomaly. Nevertheless, it is interesting to see what the predictions of LQC are. 

\begin{figure}
    \centering
    \includegraphics[width=0.8\textwidth]{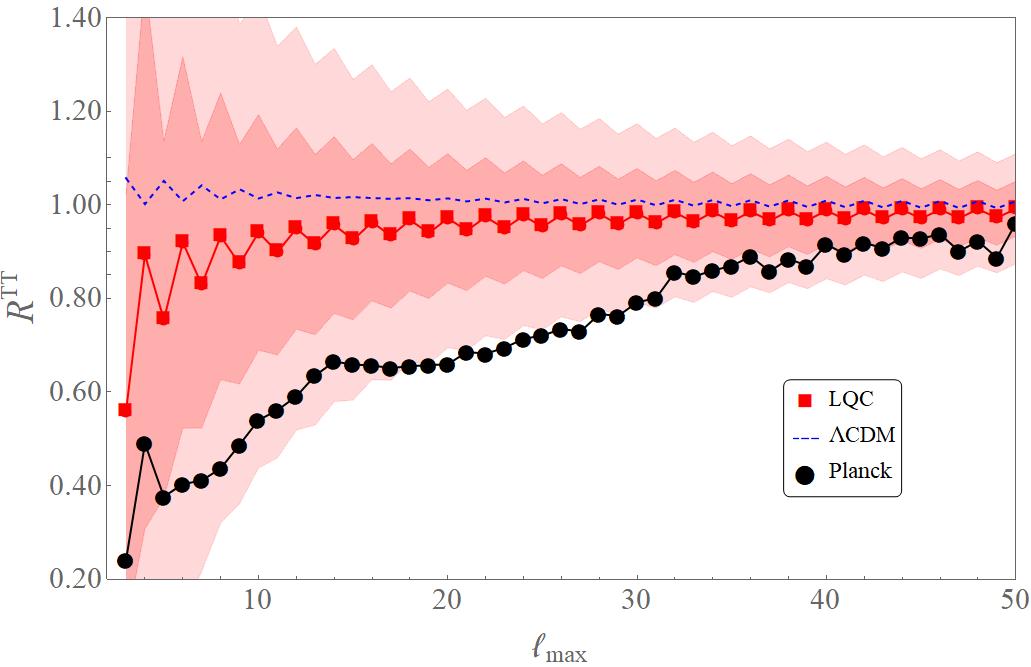}
    \caption{$R^{TT}(\ell_{\rm max})$ for the modulated spectrum generated in LQC (solid red). 
    $R^{TT}(\ell_{\rm max})$ predicted in LQC shows a preference for odd parity for low multipoles, unlike the one in the standard model (dashed, blue). }
    \label{fig:parity_alleviation}
\end{figure}

We find that the monopolar modulation induces also a preference for odd parity multipoles $\ell$, in agreement with observations. After inspection, this fact is not surprising, and it is a consequence of the simple fact that, in a power suppressed angular power spectrum, 
the sum of $\frac{\ell (\ell\,+\,1)}{2\,\pi}\, C^{\rm mod}_\ell$ starting from $\ell=2$ is larger for  odd multipoles, precisely because the sum starts at an even multipole---it would have been otherwise if the sum starts at $\ell=1$. Therefore, we find that in LQC there is a preference for odd-parity multipoles $\ell$, as measured by $R^{TT}(\ell_{\rm max})$, and the result is a consequence of the power suppression. We report our result in Fig. \ref{fig:parity_alleviation}. For comparison, we provide the corresponding values obtained in the $\Lambda$CDM model and the observations made by Planck. Although the result for $R^{TT}(\ell_{\rm max})$ from LQC is closer to the data, the  value of $R^{TT}(\ell_{\rm max})$ observed by Planck is  smaller than what we find in LQC, but the significance of the deviation is modest. In the absence of a better estimator  for the parity anomaly, it is not possible for us to make a more precise comparison.

\begin{figure}
    \centering
    \includegraphics[width=0.8\textwidth]{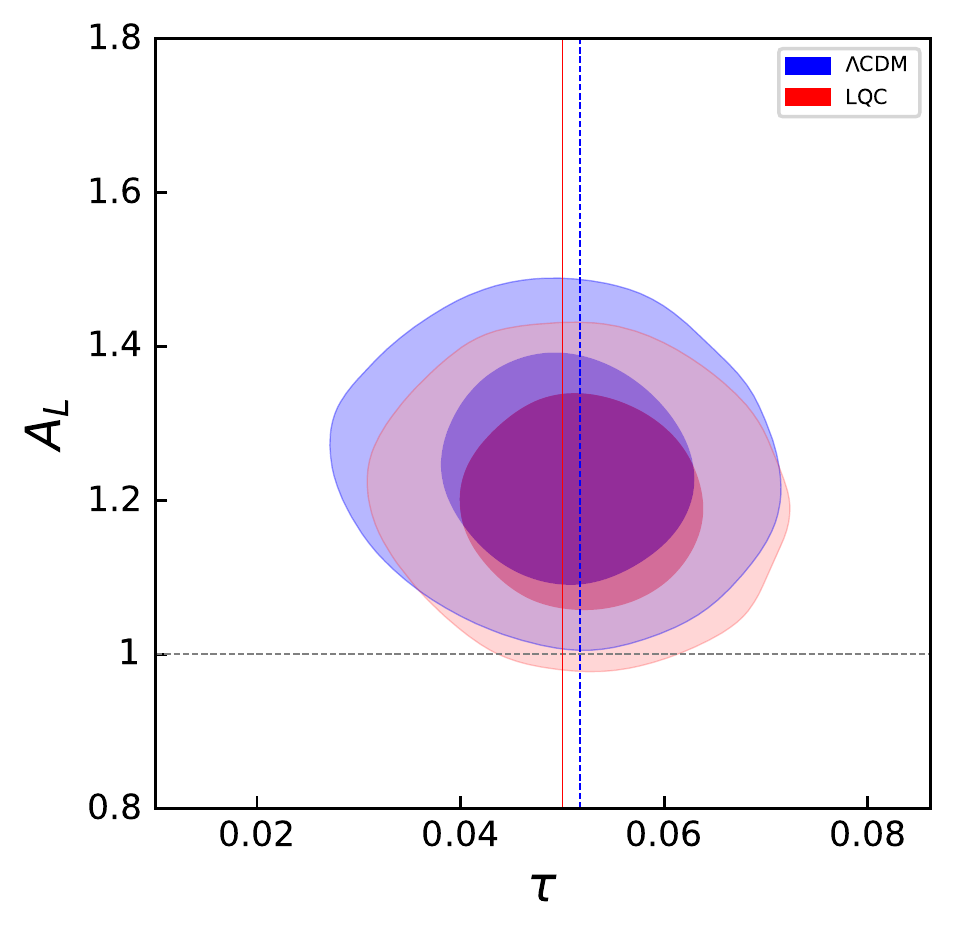}
    \caption{Marginalised joint probability distribution of $\tau$ and $A_L$ obtained from MCMC simulation for the standard model and modulated LQC. As we can see, $A_L\,=\,1$ lies within the $2-\sigma$ contour for the modulated model, thus bringing the lensing parameter closer to one.}
    \label{fig:tau_AL}
\end{figure}

\par
Yet another effect of the power suppression caused by the monopolar modulation is the alleviation of lensing tension. The relation between a power suppression and the lensing anomaly was discussed in \cite{Ashtekar:2020gec}, also in the context of LQC, and our analysis confirms the relation. The value of $A_L$  is obtained from data by  performing  MCMC simulations involving the standard six free parameters,  together with the lensing amplitude $A_L$. We repeat the analysis with the modified probability distribution obtained from LQC, using $ TT + low E$ data, and find that the marginalized mean value of the lensing parameter is $A_L\, =\, 1.20\, \pm\, 0.092$. This value is $3.5\%$ smaller than the result obtained from $\Lambda$CDM. This is a modest change. However, as shown in Fig. \ref{fig:tau_AL}, the joint probability distribution of $\tau$-$A_L$, with $\tau$ the optical depth,  shows that the value of $A_L\,=\,1$ is within $2$ standard deviations, and it is in this sense that the anomaly is alleviated.  It should also be noted, however,  that the marginalized mean value of $\chi^2$ is larger for the modulated model by $\Delta\chi^2\,=\,5.29$. This lower value of the lensing parameter $A_L$ can be explained due to the slightly larger value of $\tau$. This is because a larger value of $\tau$ implies a slightly larger value of the scalar amplitude $A_s$, which in turn leads to a smaller value of $A_L$ \cite{Ashtekar:2020gec}.

\section{Discussion}\label{sec:6}
The success of any theory  seeking to describe the unknown rests on two criteria: it should be consistent with known facts and at the same time be able to make  new predictions.  Loop quantum cosmology, as an effort to extend the $\Lambda$CDM model to the Planck regime, has met the first criteria since, when combined with inflation, it is able to produce a nearly scale invariant power spectrum and bispectrum for almost all scales in the CMB. As far as the second aspect is concerned, LQC predicts that,  
if we consider adiabatic initial conditions for perturbations  before the bounce, the primordial power spectrum and bispectrum deviate from scale invariance at wavenumbers $k\,\lesssim\, k_{LQC}$. The question is  whether these features  occur at scales that are observable today. If this is the case, then we may keep the hope to use the observations to verify LQC, and use them to further refine the theory. It is with this second aspect in mind that we investigate the link between enhanced and scale dependent perturbations generated in LQC and the CMB anomalies. 
\par
CMB anomalies, as we discussed in section \ref{sec:2}, include several features that have been observed, mostly at large angular scales in the CMB. The genuineness of these features is not under dispute. However, if considered individually, these features only depart from the predictions of standard model by p-values of $<1\%$.  Hence the possibility that some of these features appear in the CMB in a universe governed by standard $\Lambda$CDM model is not completely negligible.  However, the fact that all these seemingly distinct features occur together in our universe,  means that we either live in a  rare realization of the probability distribution of the  $\Lambda$CDM model,  or that they are signatures of  new physics. We have explored in this paper the second possibility in the context of LQC.  

In this scenario, the role of the cosmic bounce is simply to modify the initial conditions from which inflation and the $\Lambda$CDM model take over. The most relevant aspect  comes from the fact that the bounce generates strong correlations between the longest wavelengths we can observe in the CMB and longer, super-Horizon perturbations. These correlations, although cannot be observed directly in the CMB because they involve at least one super-Horizon mode,  bias the form of the power spectrum. 
This bias translates in higher probability for certain features to be realized in our CMB. We find interesting that such effect can simultaneously produce a suppression and a dipolar modulation in the sky, compatible with observations. These two features were thought to be unrelated, and LQC provides a common origin for them. It is important to keep in mind that the origin of the anomalies is probabilistic,  
and the way LQC can account for them  is by modifying the probability distribution. For instance, the dipole asymmetry does not arise in LQC as the result of breaking isotropy at the fundamental level, but rather because in a non-Gaussian universe the size of the anisotropies expected to be found by a typical observer are larger than in a Gaussian theory.

In our calculation we have adjusted a free parameter in LQC, which controls the amount of expansion accumulated from the bounce to the end of inflation. The statement is therefore that there exist a value of this parameter for which  the observed anomalies can be originated form LQC (this value is $\approx\,71$ e-folds; this number includes  the expansion during the inflationary and the pre-inflationary epochs). Our calculations also involve some approximations and limitations, and in particular we have not been able to account precisely for the effects of the oscillations in the bispectrum. It would be desirable to  investigate the way these oscillations convolve with the power spectrum and transfer functions in order to understand their effect on CMB. Furthermore, the data quantifying the  anomalies is limited, as it is based on simple estimators such as $S_{1/2}$ and the binned value of the dipolar amplitude $A_1(\ell)$. Additional data, for instance coming from tensor modes, would allow a more precise comparison of our ideas with observations. But in spite of these limitations, we find remarkable that the bounce of LQC can produce effects in the CMB which are in good consonance with the observed anomalies, regarding both the order of magnitude of the amplitudes as well as their scale dependence. The possibility that the observed features are informing us about the Planck era of the cosmos is mind-blowing, and certainly deserves further attention. Our contribution should be considered as a first step in this direction.

Finally, in this work we have assumed adiabatic initial conditions for the scalar perturbations before the bounce, wherein the unmodulated primordial power spectrum generated in LQC is enhanced at super-horizon scales. There has been a proposal in LQC \cite{Ashtekar:2016wpi,Ashtekar:2020gec} for different initial conditions, which leads to a suppressed power spectrum even before considering the non-Gaussian modulation. It would be interesting to combine both sets of ideas and compute the effect of non-Gaussian modulation in that model.

\section*{Acknowledgements}
We have benefited from discussions with A. Ashtekar, B. Bolliet, B. Gupt, J. Olmedo, J. Pullin, and P. Singh. This work is supported by the NSF CAREER grant PHY-1552603, and from the Hearne Institute for Theoretical Physics. This paper is based on observations obtained from Planck (http://www.esa.int/Planck), an ESA science mission with instruments and contributions directly funded by ESA Member States, NASA, and Canada. This research was conducted with high performance computing resources provided by Louisiana State University (http://www.hpc.lsu.edu).
\bibliography{Refs}
\end{document}